\documentclass[prl,reprint,superscriptaddress,longbibliography]{revtex4-1}
\usepackage[colorlinks=true,allcolors=blue,breaklinks=true]{hyperref}
\usepackage{graphicx}
\usepackage{siunitx}
\usepackage[T1]{fontenc}
\usepackage{lmodern}
\usepackage{bbding}
\usepackage{amsmath}

\begin{document}

\title{On-chip quantum confinement refrigeration overcoming electron-phonon heat leaks}

\author{S.~Autti}
\email{s.autti@lancaster.ac.uk}
\affiliation{Department of Physics, Lancaster University, Lancaster, LA1 4YB, UK.}

\author{J.~R.~Prance}
\affiliation{Department of Physics, Lancaster University, Lancaster, LA1 4YB, UK.}

\author{M.~Prunnila}
\affiliation{VTT Technical Research Centre of Finland Ltd, P.O. Box 1000, 02044 VTT, Espoo, Finland.}

\date{\today}

\begin{abstract}
Circuit-based quantum devices rely on keeping electrons at millikelvin temperatures. Improved coherence and sensitivity as well as discovering new physical phenomena motivate pursuing ever lower electron temperatures, accessible using on-chip cooling techniques. Here we show that a two-dimensional electron gas (2DEG), manipulated using gate voltages, works as an on-chip heat sink only limited by a fundamental phonon heat-leak. A single-shot 2DEG cooler can reduce the electron temperature by a factor of two with a hold time up to a second. Integrating an array of such coolers to obtain continuous cooldown in will allow reaching down to microkelvin device temperatures.
\end{abstract}

\maketitle

Solid-state quantum devices are typically operated at the lowest accessible temperatures, down to around 10\,mK, because the physical processes that these devices are based on vanish or weaken at higher temperatures. Lowering the temperature further is an obvious strategy for new discoveries and improved device performance in applications such as quantum computing, but lowering the temperature of electrons inside the device faces a fundamental barrier due to thermal decoupling from the refrigerator. The lowest device electron temperatures that can be reached by external heat sinking is typically $\approx \SI{4}{\milli\kelvin}$~\cite{Xia2000,Samkharadze2011,Bradley2016,jones2020progress}. Below \SI{4}{\milli\kelvin} the thermal link from electrons via phonons to the refrigerator sample holder becomes vanishingly small. Here, the tiniest heat leak entering the system, for example via measurement wires or due to external radiation, can prevent further cooldown even if the heat sink temperature approaches the absolute zero. Immersing the device and the wiring in a superfluid \cite{Levitin2022} allows lowering the electron temperature to below one millikelvin, but this only moves the decoupling to a somewhat lower temperature.

The low-temperature thermal bottleneck can be circumvented by building an on-chip cooldown cycle. The first generation devices were made by placing miniaturized demagnetization refrigerators directly on chip \cite{PhysRevLett.131.077001,Bradley2017,Palma2017a,Yurttaguel2019}, providing access deep into the microkelvin regime. This technique can only be applied to devices that tolerate magnetic fields $\sim 100\;\mathrm{mT}$ and above, and it relies on a custom refrigerator platform and special device design and fabrication, such as electroplated refrigerant on the device. In this Letter we study theoretically an electric field equivalent of on-chip adiabatic demagnetization. The occupation of sub-bands in a two-dimensional electron gas can be controlled by manipulating the profile of the confinement in one spatial dimension by the potentials of electrostatic gate electrodes (Fig. 1). The heat capacity in this system only depends on the number of occupied bands, which means that an adiabatic expansion of the electron population to a larger number of bands leads to a temperature decrease, providing an integrated and scalable on-chip electronic refrigeration method. We study this cooling effect against the fundamental phonon heat-leak in a Si-based 2DEG.  

\begin{figure}[htb!]
\includegraphics[width=1\linewidth]{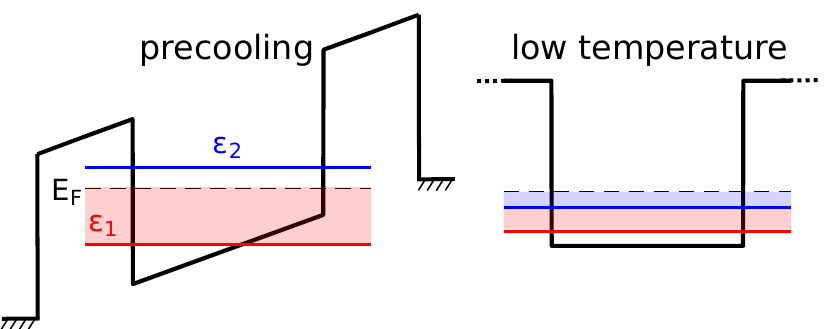}%
\caption{Schematic illustration of quantum confinement cooling using a 2DEG. During precooling, a gate voltage is applied across the confined dimension of the 2DEG to create a triangular potential (thick black line). This increases the spacing between bands corresponding to $\epsilon_1$ and $\epsilon_2$ so that only $\epsilon_1$ is below the Fermi energy. During the cooldown, the gate voltage is reduced to zero so that both $\epsilon_1, \epsilon_2 <E_\mathrm{F}$. The (partially overlapping) red and blue shaded areas are proportional to the population of each sub-band. The tilt of the potential and the level spacing are exaggerated for illustrational purposes. \label{fig:schematic}}
\end{figure}

The 2DEG heat capacity per unit area is $C= \pi^{2} k^2_\mathrm{B} T D_0(E_\mathrm{F})/3$ \cite{li1989calculated,GRIVEI199890}, where $k_\mathrm{B}$ is the Boltzmann constant, $E_\mathrm{F}$ is the Fermi energy, and $D_0(E_\mathrm{F})$ is the density of states for one 2DEG sub-band. In the absence of changes in $D_0(E_\mathrm{F})$, the 2DEG electron temperature follows

\begin{equation}\label{eq:diff_eq}
 \frac{\mathrm{d}T_\mathrm{e}}{\mathrm{d}t} = (P_0 - P_\mathrm{ep})/C.
\end{equation}
where $P_\mathrm{ep}$ is the heat flow from 2DEG electrons to phonons in the underlying material and $P_0$ contains spurious heat leaks to the 2DEG.

The entropy of an electron gas can be obtained by integrating the ratio of its heat capacity and temperature, $C(T)/T$, yielding

\begin{equation}\label{eq:entropy2}
    S(T)=\pi^{2} k^2_\mathrm{B} T D(E_\mathrm{F})/3,
\end{equation}
where $D(E_\mathrm{F})$ is the sum of the densities of state of the occupied sub-bands. Therefore, $T\propto 1/D(E_\mathrm{F})$ in any isentropic process \cite{rego1999electrostatic}.   

Let us consider the lowest two sub-bands of an isolated 2DEG at temperature $T_1\ll E_\mathrm{F}/k_\mathrm{B}$. The sub-bands have initial energies $\epsilon_0$ and $\epsilon_1$ such that $\epsilon_0<E_\mathrm{F}<\epsilon_1$; only the lowest sub-band is populated. We assume that $\epsilon_1 - \epsilon_0$ can be manipulated adiabatically \cite{prunnila2006two,zhao2022electron} by changing top and back gate voltages that control the confinement profile (Fig.~\ref{fig:schematic}). If we change $\epsilon_1 - \epsilon_0$ so that $\epsilon_0 < \epsilon_1 < E_\mathrm{F}$, the total density of available states becomes $D=2D_0$ and the 2DEG temperature is halved. Note that $E_\mathrm{F}$ will decrease during this process if the electron density is held constant. In a system with $N$ sub-bands, $N-1$ initially above $E_\mathrm{F}$, the final temperature after bringing all of them inside the Fermi surface is $T=T_1/N$. It has been shown in experiments that at least three sub-bands ($N=3$) can be manipulated this way \cite{yang2020three}. Continuous cooling can be achieved by operating multiple 2DEGs in turn to pump heat from a common bath. This is possible as long as each individual 2DEG is able to hold a reduced temperature for longer than the time taken to cool it, and to switch a thermal connections to the common bath.

\begin{figure*}
\includegraphics[width=1\linewidth]{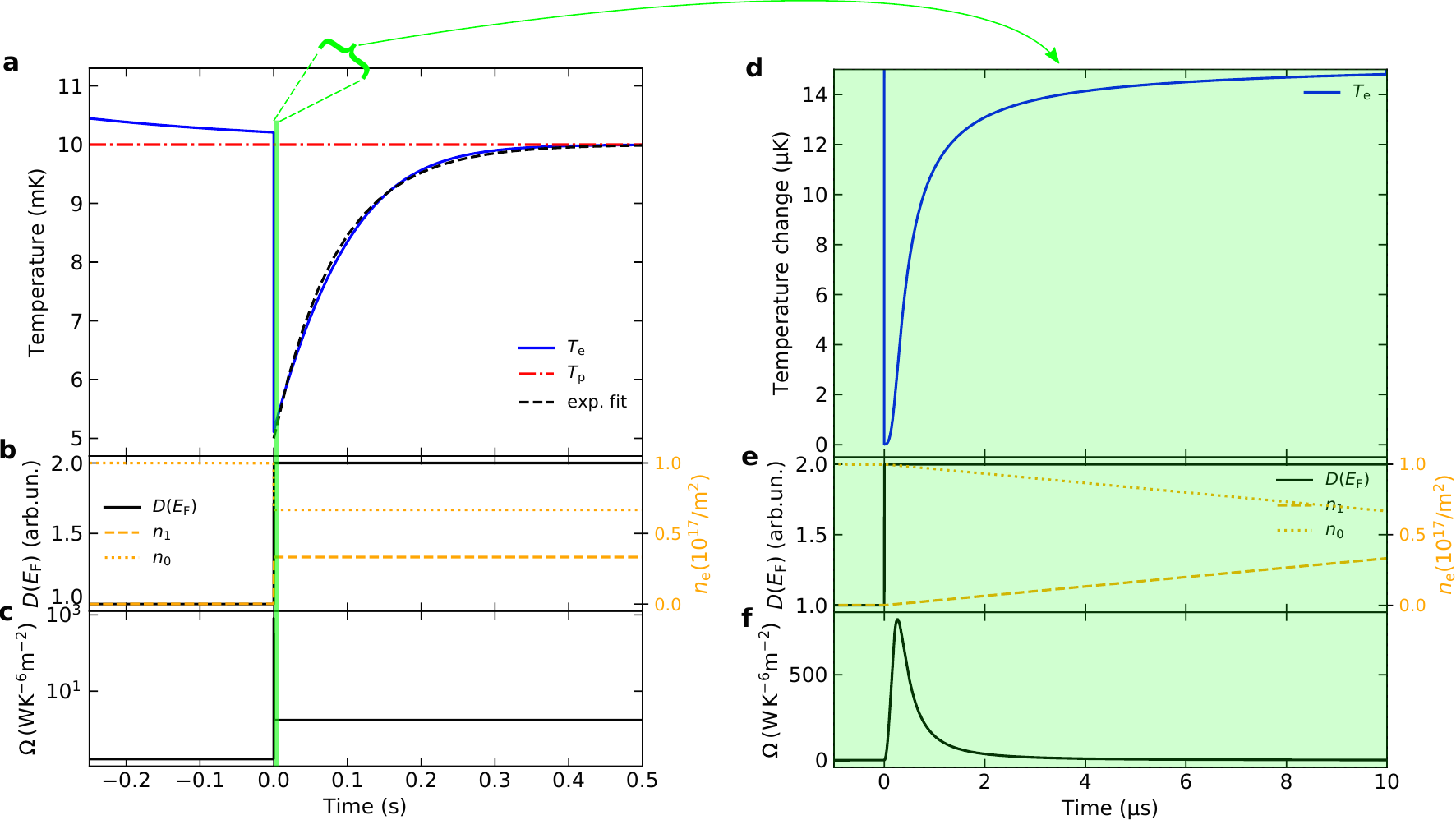}%
\caption{Simulated 2DEG cooldown performance: At $t<0$\,s only the lowest-energy band is populated ($n_1=10^{13}$\SI{}{\centi\metre^{-2}}, $n_2=0$). The spurious heat leak is set to $P_0=0$ for simplicity. Thus, the simulated electron temperature shown in panel ({\bf a}) (blue line) comes to equilibrium at $T_\mathrm{e} = T_\mathrm{p}$. At $t=0$, one third of the electrons are moved to the second-lowest band in a linear ramp that lasts \SI{10}{\micro\second} (right y axis in panel ({\bf b})). During this process, the heat leaks are negligible and the density of states $D(E_\mathrm{F})$, shown on the left y axis in panel ({\bf b}), is doubled, corresponding to bringing a second 2DEG sub-band inside the Fermi sphere. Thus, $T_\mathrm{e}$ is halved, after which it returns to an equilibrium value approximately exponentially (black dash line) with the time constant $\tau=90$\,ms. The electron-phonon coupling increases in this process (panel ({\bf c})). A zoomed-in version of the population ramp showing the transient in detail is in panels ({\bf d},{\bf e},{\bf f}). When the electrons start to move from $n_0$ to $n_1$ at $t=0$\,s, the coupling becomes momentarily very large (panel ({\bf f})). The warm-up rate just after $t=0$ is large but it quickly decreases as the population ramp continues, and the temperature increase remains negligible during the transient. Before $t=0$ the electron-phonon coupling via the upper band is taken to be zero. We have assumed here that the density of states (panels ({\bf b}, {\bf d}) left y axis) increase takes place at $t=0$, meaning that the heat capacity increase and thus the temperature decrease take occur instantaneously at $t=0$. In a realistic experiment, the cooldown may be smoothed for example due to spatial inhomogeneities of the 2DEG electron density, but for an adiabatic process this does not change the cooldown performance, only the details during the transient.\label{fig:depol_performance1}}
\end{figure*}

In experiments, the 2DEG temperature evolves adiabatically provided the changes in the sub-band population are slow as compared with the inter-band coupling, typically reported to be of the order of picoseconds \cite{casse2020evidence,lee1999intersubband}. This condition should be always satisfied for gate operations that last a \SI{}{\micro\second} or more. The cooldown performance is primarily hindered by heat leaks from the environment. We consider a 2DEG in silicon placed in a commercial dilution refrigerator that cools the phonons in the underlying semiconductor to $T_\mathrm{p}=10$\,mK \cite{PhysRevLett.131.077001}.  Where relevant, the 2DEG surface area is taken to be $A=$\SI{100}{\micro \meter}$\times$\SI{100}{\micro \meter} and the thickness $d=100$\,nm.

Phonons can carry heat across several material interfaces and remain at the refrigerator temperature throughout the on-chip cooldown process \cite{PhysRevLett.131.077001}. The electron-phonon coupling decreases rapidly at low millikelvin temperatures, allowing the on-chip electrons to reach temperatures below the phonon temperature. We write the electron-phonon heat flow per unit area as $P_\mathrm{ep}=\Omega (T_\mathrm{e}^6- T_\mathrm{p}^6)$, where the coupling for a single sub-band can be written as \cite{prunnila2007electron}

\begin{equation}\label{eq:theory_ep_full}
   \Omega = \frac{\hbar \protect D_0 }{2\protect\pi ^{2}\protect\rho\kappa }B_{5}\frac{1}{\kappa l_s}\left\langle \frac{\sin ^{2}\theta
\Xi _{S}^{2}}{(l_\mathrm{s}\kappa )^{-2}+\sin ^{2}\theta }\right\rangle \left( \frac{k_{B}}{\hbar v_\mathrm{s}} \right)^{6}\;.
\end{equation}
In this expression $v_\mathrm{s}$ is the sound velocity, $\rho$ is the mass density of the material, $v_\mathrm{F}$ is the Fermi velocity, $B_5\approx122$, $l_\mathrm{e}$ is the electron mean free path, the length scale $l_\mathrm{s}=l_\mathrm{e}v_\mathrm{F}/(2 v_\mathrm{s})$, $\kappa$ is the Thomas-Fermi screening wave vector, and $\Xi _{S}$ is the angle-dependent deformation potential. Electron mean free path is defined by electron mobility $\mu$ through $l_\mathrm{e} = v_\mathrm{F} \mu m^{*}/e$, where $m^{*}$ is the effective mass. Values and expressions needed for evaluating Eq. (\ref{eq:theory_ep_full}) are given in Table~\ref{table:parameter_values} in the Appendix. Eq. (\ref{eq:theory_ep_full}) applies in the diffusive limit of electron-phonon coupling, where thermal phonon wavelength $q_{T}=\frac{k_{B} T_\mathrm{p}}{\hbar v_{s}}$ exceeds $l_\mathrm{e}$. The pre-factor with average over a solid angle that appears in the Equation can be approximated as

\begin{equation}    \left\langle \frac{\sin ^{2}\theta
\Xi_{S}^{2}}{(l_\mathrm{s}\kappa )^{-2}+\sin ^{2}\theta }\right\rangle  \approx \Xi^{2} \left( 1-\frac{1}{\sqrt{(l_\mathrm{s} \kappa)^2+1}} \right),
\end{equation}
where $\Xi$ is a ``scalar'' deformation potential constant in the range 1-10 eV. We use $\Xi=10$\,eV in the below simulation as a pessimistic estimate \cite{hull1999properties}.  Note that thanks to the solid-angle average pre-factor of $F(T)$ of Ref.~\cite{prunnila2007electron} there is no unphysical divergence of F(T) at $n_e \longrightarrow 0$ unlike in the formulae of Ref. \cite{Sergeev2005}.

\begin{figure}[htb!]
\includegraphics[width=1\linewidth]{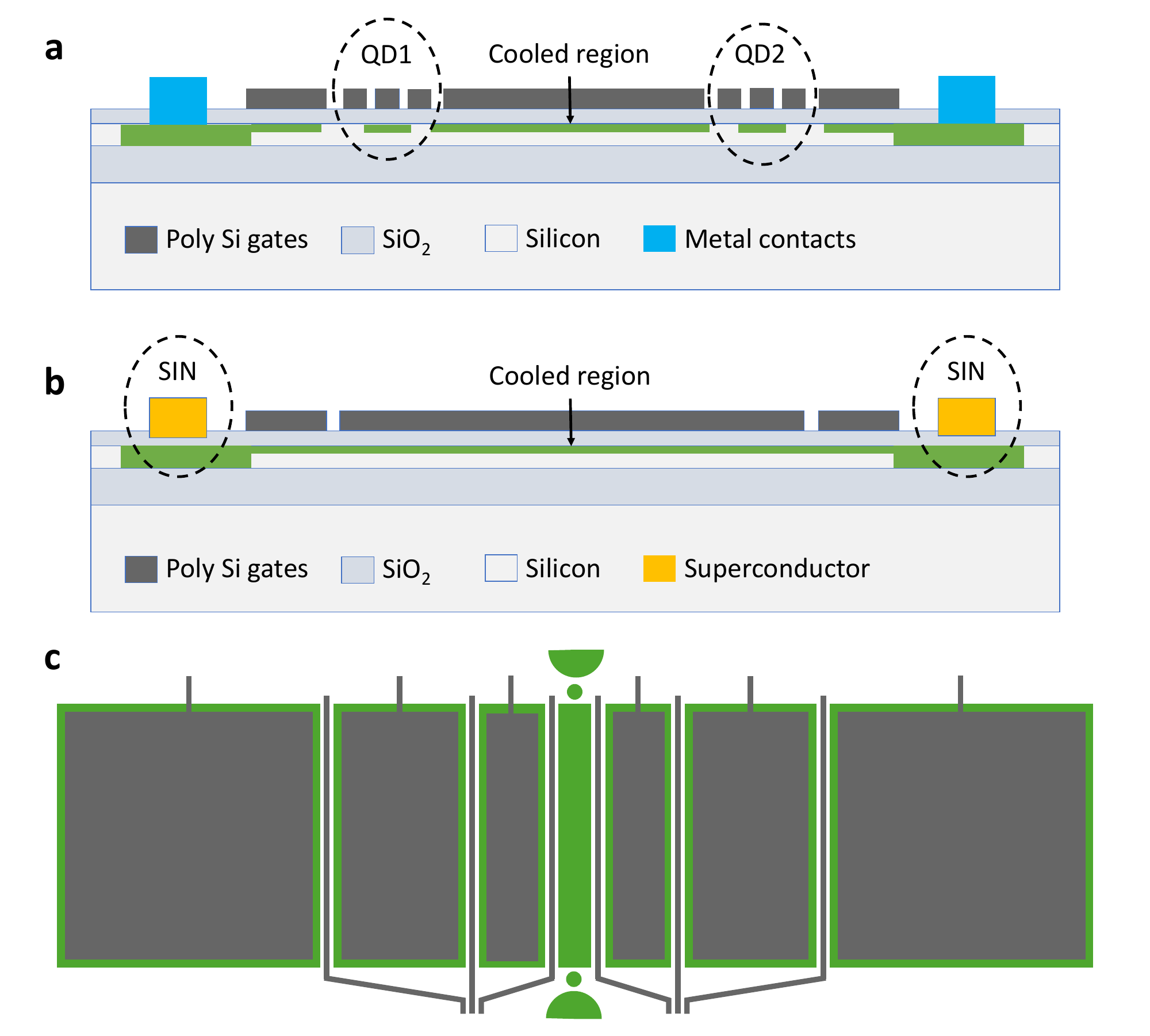}%
\caption{Practical schemes for implementing 2DEG cooling and temperature measurement. The 2DEG is created in a silicon layer in silicon oxide using patterned polysilicon gates. Conducting regions are shown in green. In design ({\bf a}), shown in cross section, two quantum dots (QD1 and QD2) are connected to the 2DEG via tunnel barriers. The 2DEG is cooled by changing the voltages of the polysilicon gate above it and a bulk silicon back gate below. The temperature of the cooled 2DEG can be found from transport measurements through the 2DEG via the quantum dots as in Ref. \cite{Prance2009}. In the design ({\bf b}) two superconductor-insulator-normal metal (SIN) junctions are connected to the 2DEG. Here the N region below the S electrode is formed by degenerately doped silicon in the spirit of Ref. \cite{Mykkaenen2020}. The electron temperature is inferred from transport measurements through the SINIS structure. In design ({\bf c}), shown from above, two chains of cooled 2DEGs (green), each containing three 2DEG regions in series, are connected in parallel to a central cooled region. Electrical and thermal connection between the 2DEG regions is controlled by narrow polysilicon gates (gray) between them. Sequential cooling of the regions connected in series can be used to reach a lower final temperature. Alternating operation of the parallel chains can be used to achieve continuous cooling of the central region.\label{fig:EDR_coolers}}
\end{figure}

The 2DEG cooldown simulation shown in Fig.~\ref{fig:depol_performance1} starts from an elevated $T_\mathrm{e}$, reaching a stable value exponentially. The population of the lowest band is initially $n_1=10^{13}\,$\SI{}{cm^{-2}} and that of the second sub-band $n_2=0$. The 2DEG needs to be well isolated electrically to prevent the Wiedemann-Franz heat leak from rapidly overwhelming the electron heat capacity (see Fig.~\ref{fig:EDR_coolers}c). For simplicity, $P_0=0$ in this calculation, and therefore $T_\mathrm{p}=T_\mathrm{e}=10$\,mK in equilibrium. 

At $t=0$, we move a third of the electron population to the upper band in a linear sweep that lasts \SI{10}{\micro\second}. We label this operation a ``transient''. In experiments, the timescale of the transient is limited from below by the charging time of the top gate whose capacitance is $\lesssim 10\,\mathrm{pF}$
and the charging time can be as low as $\sim 1\,\mathrm{ns}$ with a low-resistance connection ($\sim 10\,\Omega$). 

With two bands populated, the total electron phonon coupling is given by the sum of the couplings calculated separately for the two sub-bands using Eq.~(\ref{eq:theory_ep_full}). We assume that the density of states is doubled immediately as the electrons start to move (this assumption is justified below). The imposed temperature reduction is much faster than the dynamics described by Eq.~(\ref{eq:diff_eq}), causing a discrete jump in the simulated $T_\mathrm{e}$. Consequentially, the electron temperature is initially halved to $T_\mathrm{e}=5$\,mK and it then returns exponentially to the equilibrium value with the time constant $\tau\approx0.1$\,s.

The electron-phonon coupling in Eq.~(\ref{eq:theory_ep_full}) depends on sub-band electron density. For typical parameter values it has a maximum between $n_\mathrm{e}=0$ and $n_\mathrm{e}=10^{12}\,$\SI{}{\centi\metre^{-2}}. Most of the parameters in Eq.~(\ref{eq:theory_ep_full}) have simple dependencies on electron density provided in Table~\ref{table:parameter_values}, but how the electron mean-free-path  depends on the sub-band electron density is more subtle. For a single occupied sub-band, as the total electron density decreases, the charge carrier elastic scattering time and mobility - and, thereby, the mean-free-path - typically decrease below the so-called peak mobility defining the crossover between Coulomb and surface roughness scattering \cite{ando1982electronic}. To capture the essential features of the mean-free-path and mobility density dependence we assume that the mobility for the upper band decreases linearly to zero as a function of electron density below the final value of $n_2$, and above this value the mobility is the same for both bands ($\mu=10^4$\,\SI{}{\centi \metre^2 \, \volt^{-1} \, \second^{-1}}). 

We now concentrate on the transient shown in the right panels of Fig.~\ref{fig:depol_performance1}. The coupling increases by an order of magnitude during the transient, but the electron temperature only increases by a few \SI{}{\micro\kelvin} before the transient ends. This conclusion holds even if we fix the mobility (mean free path) to the value that corresponds to the maximum coupling, reached at some point during the transient. Therefore, the transient can safely be ignored, and the cooldown performance remains largely the same regardless of the precise timing of the increase in the density of states, changes in electron mobility during the transient, and also regardless of inhomogeneities in the system that mean the population transfer between the 2DEG bands does not take place uniformly with-in the 2DEG area.

In addition to the heat flow between electrons and phonons, experiments at microkelvin and millikelvin temperatures typically observe a spurious heat leak $P_0$, originating from noise entering via measurement contacts, a slowly decaying intrinsic relaxation in dielectrics \cite{PhysRevLett.131.077001,heikkinen2014microkelvin,hosio2011propagation} or possibly from external radiation such as cosmic rays \cite{QUEST, QUEST2}. 
It is known that that $P_0$, scaled to the thickness of the 2DEG system considered here, can be 1\,aW/($10^4$\SI{}{\micro\meter})$^2$ \cite{PhysRevLett.131.077001} in structures in a magnetic field. We consider this as the worst case scenario. With this large heat leak, the dynamic steady state temperature of electrons is $T_\mathrm{e}\approx 30$\,mK (when electron-phonon coupling is absorbing $P_0$) for $T_\mathrm{p}=10$\,mK. The cooldown time constant is reduced from $\sim 0.1$\,s to $\sim 10$\,ms. 

In the absence of a magnetic field, $P_0$ can be much lower. Transport measurements across an interface between two dielectric materials in the zero-temperature limit yield $P_0 \sim$\,zW/($10^4$\SI{}{\micro\meter})$^2$ \cite{heikkinen2014microkelvin,hosio2011propagation}.  Thus, we also consider a well isolated 2DEG where we find $P_0$ by fitting it so the initial temperature is $T_\mathrm{e}=15$\,mK, yielding $P_0\approx 20$\,zW (still pessimistic as compared with the best results reported in literature \cite{Levitin2022}). With this heat leak the cooldown time constant is already similar to that obtained with $P_0=0$. We emphasise that in a realistic best-case scenario \cite{Levitin2022} the electrons would be cooled close to 1\,mK before starting the on-chip cooldown cycle, with the phonon temperature yet lower than this. Here, $\tau$ becomes orders of magnitude longer than a second.

We can also try to simulate the 2DEG cooldown performance based on experimental measurements of electron-phonon coupling in silicon and GaAs 2DEGs \cite{PhysRevLett.88.016801,mittal1996electron,proskuryakov2007energy,appleyard1998thermometer,prunnila2005intervalley}. This effort is hindered by the unavailability of measurements of the electron-phonon coupling below 100\,mK, but in the absence of more applicable experimental data we extrapolate them to 10\,mK. These measurements have also been carried out at electron densities around $n_\mathrm{e}=10^{11}$\SI{}{\centi \metre^{-2}}, where the electron-phonon coupling is two orders of magnitude stronger than at $n_\mathrm{e}=10^{13}$\SI{}{\centi \metre^{-2}}. Keeping these shortcomings in mind, and without reducing the coupling at higher electron densities in the simulation, the obtained cooldown time constants \cite{PhysRevLett.88.016801,mittal1996electron,proskuryakov2007energy,appleyard1998thermometer,prunnila2005intervalley} range between $\sim 10$\,ms and 1\,s, in good agreement with the purely theoretical values.

We propose two schemes for measuring the performance of the 2DEG refrigeration scheme (Fig.~\ref{fig:EDR_coolers}a,b). Manipulation of the 2DEG sub-band occupation requires gates above and below the electron gas. Both devices shown in Fig.~\ref{fig:EDR_coolers} are based on a 2DEG in an intrinsic SOI structure \cite{prunnila2006two,zhao2022electron}, with a bulk back-gate and a polysilicon top-gate to adjust the sub-bands as in Fig. \ref{fig:schematic}. The cooled region of 2DEG is isolated from external electrical systems by the in-plane shape of the polysilicon gates. The temperature of the cooled region could be measured using either quantum dots (Fig.~\ref{fig:EDR_coolers}a) or superconducting junctions (Fig.~\ref{fig:EDR_coolers}b). Both types of device allow the temperature of the isolated region to be measured without introducing an additional heat leaks; indeed they could be operated to provide even some background cooling\cite{Prance2009,Feshchenko2014,Mykkaenen2020}. 

Figure~\ref{fig:EDR_coolers}c shows how multiple cooled 2DEG regions could be operated in parallel and in series to reach lower temperatures and achieve continuous cooling. Such schemes are possible because the thermal connection between different 2DEG regions can be easily controlled using additional gates to electrically connect or isolate the regions.

The schemes proposed in Fig.~\ref{fig:EDR_coolers} could serve to validate and characterize the 2DEG refrigeration as described here. They also show how to integrate 2DEG cooler with quantum dot based spin-qubits. The transient between one and two populated sub-bands is experimentally unexplored and measuring the electron-phonon coupling during this process can be a challenging task. The theoretical assumptions we have made to describe the transient seem well justified, but characterizing the proposed devices would serve as a way to explore this novel regime experimentally.

We have chosen pessimistic parameter values and operational conditions in simulating the 2DEG cooldown performance. For example, decreasing the phonon temperature to 5\,mK, which has been reported in multiple 2DEG experiments and can be achieved on commercial refrigerator platforms, decreases the minimum electron temperature in single-stage operation to 2.5\,mK and increases the time constant to $\tau=1.5$\,s. We also note that it is possible to control the population of three bands \cite{yang2020three}, not just two. This would allow reducing the lowest temperature reached to one third of the starting temperature. Finally, we note that as elastic scattering increases the electron-phonon coupling using high-mobility systems could improve the cooldown performance. Devices of interest could be based, e.g., on SiGe \cite{melnikov2015ultra,shashkin2019recent} or III-V quantum wells.

\begin{acknowledgments}
All the simulation data in this Letter are available at https://doi.org/10.17635/lancaster/researchdata/xxx, including descriptions of the data sets. The simulation codes can be obtained from the corresponding author upon reasonable request.

This research is supported by the U.K.\@ EPSRC (EP/W015730/1), the European Union's Horizon 2020 research and innovation programme  (European Microkelvin Platform 824109 and EFINED 766853) and EIC Transition programme (SoCool 101113086), by the Academy of Finland through the Centre of Excellence program (projects 336817 and 312294), and by Business Finland through QuTI-project (40562/31/2020).
\end{acknowledgments}

%

\begin{widetext}

  \clearpage  
\section{Appendix}

\setcounter{figure}{0}
\setcounter{table}{0}

\renewcommand{\theequation}{A\arabic{equation}}
\renewcommand{\thefigure}{A\arabic{figure}}
\renewcommand{\thetable}{A\arabic{table}}

\begin{table*}[b!]
  \centering
  \caption{Theoretical simulation parameters for a silicon 2DEG used in the simulation in Fig.~\Ref{fig:depol_performance1}, and the expressions used to obtain their values. Where a value is given instead of an expression, these are measured values for silicon obtained from the cited sources. Here $\Gamma(n)$ is the gamma function, $\zeta(n)$ is the Riemann zeta function, $v_\mathrm{T}$ stands for transverse sound velocity, and $v_\mathrm{L}$ for longitudinal sound velocity, .}
  \vspace{0.3cm}
  
  \begin{tabular}{c|c|c|c|c|c}
    
       Parameter & Symbol & Expression & Value & Material & Ref.  \\
    \hline 
    Electr. 2D density of states & $D_0(E_\mathrm{F})$ & $D_0(E_\mathrm{F}) = \eta  m^{*}/(2 \hbar^2 \pi)$ & & & \\
    Electr. in-plane eff. mass & $m^{*}$ &  & $0.19\, m_\mathrm{e}$  & Si & \\
    sub-band degeneracy & $\eta$ &  & 4 & & \\    
   
    Electron density & $n_e$ &  & $10^{15} - 10^{17}\, \mathrm{m^{-2}}$ & Si &  \\
    Fermi energy & $E_{F}$ & $E_F = n_e/ D_0$ & & &  \\
    Fermi velocity & $v_\mathrm{F}$ & $v_\mathrm{F} = \sqrt{2 E_F /m^{*}}$ &  & &  \\
    Electron mobility & $\mu$ &  & $\sim 10^4\,\mathrm{cm^2/V/s}$ & & \cite{muhonen2011strain,hull1999properties} \\
    Electron mean free time & $\tau$ & $\tau=\mu m^{*} /e$ & & &  \\
    Electron mean free path & $l_\mathrm{e}$ & $l_\mathrm{e}=v_F \tau=\sqrt{2 \pi n_e/\eta}\, \hbar \mu /e$ & & &  \\
    Screening wave vect. (Th.-F.) & $\kappa$ & $\kappa =D_0 e^{2}/(2\varepsilon _{b})$ & & &  \\
    Bare electric permittivity & $\varepsilon _{b}$ & & $11.7 \varepsilon_{0}$ \ & & \cite{cardarelli2008materials} \\
    Sound velocity & $v_\mathrm{s}$ & $v_s\approx \min(v_\mathrm{T},v_{L}) $ & $ 4.7\cdot 10^3$\,m/s & Si & \cite{muhonen2011strain,hull1999properties} \\
    Mass density of the crystal & $\rho$ &  & $2.3\, \mathrm{kg/m^3}$ & Si & \cite{muhonen2011strain,hull1999properties} \\
    Scalar deformation potential & $\Xi$ &  & $\sim 10$\,eV & Si & \cite{hull1999properties,muhonen2011strain} \\
    & $B_5$ & $B_{n-1} = \Gamma(n)\zeta(n)$  & $8 \pi^6/63 \approx 122$ & & \cite{prunnila2007electron} \\
  \end{tabular}\label{table:parameter_values}
\end{table*}

\end{widetext}

\end{document}